\documentclass[conference]{IEEEtran}
\usepackage{cite}
\usepackage{url}

\ifCLASSINFOpdf
   \usepackage[pdftex]{graphicx}
\else
\fi
\usepackage{multirow}
\usepackage{multicol}  
\usepackage{array}
\usepackage{balance}
\usepackage{microtype}
\usepackage[all]{nowidow}

\newcolumntype{L}[1]{>{\raggedright\let\newline\\\arraybackslash\hspace{0pt}}m{#1}}
\newcolumntype{C}[1]{>{\centering\let\newline\\\arraybackslash\hspace{0pt}}m{#1}}
\newcolumntype{R}[1]{>{\raggedleft\let\newline\\\arraybackslash\hspace{0pt}}m{#1}}

\usepackage{amsmath}
\usepackage{algorithm}
\usepackage{algpseudocode}
\algrenewcommand{\algorithmiccomment}[1]{{\color{mygreen}//#1}}
\algrenewcommand\algorithmicprocedure{\textbf{kernel}}

\usepackage{fancyvrb}
\usepackage{xcolor}
\definecolor{mygreen}{RGB}{0, 102, 0}
\definecolor{myblue}{RGB}{6, 0, 74}

\begin{document}
%
\title{C-for-Metal: High Performance SIMD Programming on Intel GPUs
}


%
\author{\IEEEauthorblockN{Guei-Yuan Lueh,
Kaiyu Chen,
Gang Chen, 
Joel Fuentes,
Wei-Yu Chen,
Fangwen Fu,\\
Hong Jiang,
Hongzheng Li, and
Daniel Rhee}
\IEEEauthorblockA{Intel Corporation\\
Santa Clara, CA, USA\\
\{guei-yuan.lueh, kai.yu.chen, gang.y.chen, joel.fuentes, weiyu.chen, fangwen.fu,\\
hong.h.jiang, hongzheng.li, daniel.rhee\}@intel.com}}


\maketitle

\begin{abstract}
The SIMT execution model is commonly used for general GPU development. 
CUDA and OpenCL developers write scalar code that is implicitly parallelized by
compiler and hardware. On Intel GPUs, however, this abstraction has profound 
performance implications as the underlying ISA is SIMD and important 
hardware capabilities cannot be fully utilized. To close this performance
gap we introduce C-For-Metal (CM), an explicit SIMD programming
framework designed to deliver close-to-the-metal performance on Intel GPUs. 
The CM programming language and its 
vector/matrix types provide an intuitive interface to exploit the underlying 
hardware features, allowing fine-grained register management, 
SIMD size control and cross-lane data sharing. Experimental results show 
that CM applications from different domains
outperform the best-known SIMT-based OpenCL implementations,
achieving up to 2.7x speedup on the latest Intel GPU.
\end{abstract}


%
\IEEEpeerreviewmaketitle

\section{Introduction}




Mainstream GPU programming as exemplified by CUDA \cite{nickolls2008scalable} and OpenCL \cite{munshi2009opencl} employ a ``Single Instruction Multiple Threads'' (SIMT) 
programming model. The CPU host code in an OpenCL application defines an N-dimensional computation grid where each index 
represents an element of execution called a ``work-item''. 
An OpenCL kernel describes the algorithm that will be executed on GPU for one work-item.
Work-items are grouped together into independent ``work-groups'' that execute concurrently.
Work-items inside one work-group may communicate through fast on-chip shared local memory (SLM) and barrier synchronization.

OpenCL's programming model is a powerful paradigm to express data parallelism, as developers can write purely scalar code
for their kernels without knowing the details of how the work-items are mapped to the hardware execution units.
This abstraction has profound performance implications, however, as the Intel GPU architecture (also called Gen) and the underlying instruction set
architecture (ISA) is ``Single Instruction Multiple Data'' (SIMD). 
Intel GPUs feature an expressive instruction set that supports 
variable SIMD-sizes as well as powerful regioning capabilities that allow for fast cross-lane data sharing.
An execution unit (EU) on Gen has a fixed number of hardware threads, and each thread executes SIMD instructions 
on its dedicated 4KB byte-addressable register file.
The OpenCL compiler is responsible for vectorizing the kernel into one of the three SIMD sizes (8, 16, 32) 
for thread dispatch, and work-items execute the same instructions on one thread in lock-step. 
SIMD size selection is thus the most important optimization decision for the compiler, as it affects thread occupancy,
instruction-level parallelism (ILP), SIMD-lane utilization due to divergence, and register spill. 

A high-performance program on Gen needs to exploit a thread's dedicated register file to cut down memory traffic
while avoiding register spill, which is often fatal for performance.
This can be surprisingly difficult to achieve for OpenCL programs, however, as in order to stay portable 
the language offers no mechanism for direct register file control. 
Register pressure estimate at the source level is often wildly inaccurate
due to the various compiler optimizations and transformations that must happen to lower OpenCL C into Gen ISA.

Since under the SIMT model each work-item executes independently, OpenCL programs also lose control 
of data sharing among the cooperative items in the same thread. 
Furthermore, the SIMT model prevents OpenCL programs from directly accessing Gen ISA's powerful 
regioning mechanisms, which allows one SIMD lane to access another lane's data at no additional cost. 
The introduction of subgroups in OpenCL 2.0 partially alleviates the gaps by exposing some of the underlying
hardware capabilities through builtin functions, but getting close to the metal performance
with OpenCL on Intel GPUs remains challenging.

%

This paper presents the C-for-Metal (CM) development framework, an explicit SIMD programming model designed specifically for coding to the metal on Intel GPUs. The CM language is an extension to C/C++ that provides an intuitive interface to express explicit data-parallelism at a high level of abstraction. At the core of the language are two special vector and matrix types that form the foundation of its programming model. Vector and matrix variables are to be allocated in registers, which makes it much easier to control register usage at the source level. 
A CM kernel describes the algorithm for an entire hardware thread instead of a single work-item through builtin operations on vectors and matrices; of particular importance is the select operator that supports efficient register-gather of elements in a variable and is mapped directly to the Gen ISA regions. Programmers explicitly control an instruction's SIMD size by varying the number of elements returned in a select operation, and different SIMD sizes may be used based on considerations such as register demand and divergence.

The CM compiler (CMC) is based on the LLVM infrastructure \cite{lattner2004llvm} and is responsible for generating 
Gen ISA SIMD instructions from the high-level vector and matrix operations. A number of CM-specific intrinsics
are introduced to effectively represent such operations in the LLVM intermediate representation (IR). A sequence of CM-specific optimizations and transformations are developed around those intrinsics. One unique challenge in developing this compiler is that we need to strike a careful balance between compiler optimizations and What-You-Write-is-What-You-Get.
CM kernels are fully compatible with the Intel GPU OpenCL runtime \cite{neo2020} and oneAPI Level Zero \cite{oneapi} and can be launched directly as if they are written in OpenCL.
While Gen is CM's native architecture, CM kernels may also be executed on CPU for debugging purposes. The CM development framework is open source and can be found in \cite{cm2018}.

We present a comprehensive experimental evaluation of representative applications from different domains implemented in CM and OpenCL. 
For each workload we provide an implementation sketch on how to code to the metal on Gen using CM. We show that CM kernels 
achieve up to 2.7x speedup compared to the best-known OpenCL implementations that use available Intel-specific GPU extensions \cite{subgroup}. 
The speedup offered by CM does not mean a sacrifice to productivity; while OpenCL may allow for rapid prototyping of sequential code, 
this advantage is often negated by the subsequent tuning efforts required to obtain good performance on GPUs. Results from the development process of several compute kernels indicate that CM provides 2-3x more productivity in terms of the development effort than OpenCL. 

The rest of the paper is organized as follows: Section \ref{section:related-work} briefly covers the related work; Section \ref{section:motivations} discusses the main motivations of CM as an efficient SIMD programming model; Section \ref{section:programming-language} describes the CM programming language; Section \ref{section:compiler} describes the CM compiler; Section \ref{section:experimental-evaluation} presents several applications implemented in CM and their experimental evaluation; and finally Section \ref{section:conclusions} concludes this paper.

\section{Related Work}
\label{section:related-work}


SIMT and SIMD are two dominant programming models that express data parallelism.  CUDA \cite{nickolls2008scalable} and OpenCL \cite{munshi2009opencl} are two representative SIMT programming languages. 
In addition to SIMT execution,
OpenCL also supports a task parallel programming model in which a work-group contains a single work-item and parallelism is expressed via vector data types and multiple task enqueues. However, SIMT remains the dominant choice by far for OpenCL GPU implementations.

As OpenCL is designed to be cross-platform, it does not reflect the full architectural features for any specific hardware implementations. 
As a result, OpenCL is generally acknowledged to suffer from poor performance portability 
\cite{1016024, du2012cuda, pennycook2013investigation, 10.1007/978-3-642-38750-0_11}, 
and time-consuming tuning efforts including the use of non-portable vendor extensions are often mandatory to 
obtain good performance. 
Auto-tuning \cite{7284453} has long been suggested as a method to improve OpenCL's performance portability,
but given the wide disparities among the underlying hardware architecture it is unclear if 
such techniques can be generally applicable.

\cite{10.5555/2066302} presented a comprehensive performance comparison of CUDA and OpenCL and concluded 
that OpenCL programs can achieve similar performance to CUDA "under a fair comparison" once
differences in optimization strategies and compilers are accounted for. Their study is performed on NVIDIA GPUs which employ a SIMT
architecture that naturally match both CUDA and OpenCL's execution model. In contrast, CM is designed specifically 
for Intel GPUs and adopts an explicit SIMD programming model to fully exploit the Gen architecture. Most 
implementation techniques used in our CM workloads are simply not available in the OpenCL language.


SIMD programming on the CPU is conventionally done via C-style intrinsics\cite{intel-intrinsics}, 
but such assembly-like interface demands significant coding efforts.
As a result many high level SIMD programming models for C++ have been proposed. 
Together they cover a wide design spectrum from implicit vectorization (e.g., OpenMP) akin to OpenCL to 
explicit vectorization (e.g., std::experimental::simd in C++\cite{c++-simd}) similar to CM.
 \cite{10.1145/2870650.2870653} provides an evaluation of several SIMD programming models against intrinsic programming.
None of these SIMD programming models are natively designed for Gen, although a few such as OpenMP have 
been ported. More recently Intel has announced oneAPI Data Parallel C++\cite{dpc++}, 
which provides a unified, standards-based programming model for Intel architectures including CPU, GPU, FPGA, and AI accelerators.
We choose OpenCL for performance comparison as it is the most common language for general-purpose GPU programming 
on Gen and has very mature toolchain support.


CM is inspired by C* \cite{rose1987c} and VecImp \cite{leissa2012extending}. Every statement including control flow branch in VecImp is executed in a scalar or vector context explicitly. C* declares parallel variables with shape that contain many data elements. Arithmetic operators on parallel variables perform operation on all elements of a parallel variable at the same time. 


In terms of compiler infrastructure, such as LLVM, vector representations and transformations that we have explored for implementing CM are ongoing research topics. Recently, authors in \cite{lattner2020mlir} introduce MLIR, an extensible multi-level intermediate representation, which is aimed to "improve compilation for heterogeneous hardware, reducing the cost of building domain specific compilers''. MLIR community is actively working on a vector dialect. One rationale explained in \cite{mlirvector} for developing this vector dialect is ``higher-dimensional vectors are ubiquitous in modern HPC hardware''.

CM can also serve as a back-end compiler of other domain-specific languages aimed to tackle computationally expensive problems. Recent proposals for neural networks \cite{truong2016latte, rotem2018glow} and image analysis \cite{chiw2012diderot} provide high level of abstraction where the CM back-end compiler naturally fits in to target Intel GPU.

The CM language was invented more than ten years ago, and hundreds of CM applications have been developed inside and outside Intel.
As an example in \cite{fuentes2019lock} and \cite{10.1007/978-3-030-43229-4_33}, authors study the extension of linearization properties to SIMD programming using CM, including the implementation of a concurrent data structure using atomic operations. 

\section{Motivations for a New Programming Model on Gen}
\label{section:motivations}


Here we describe three main challenges faced by SIMT models as represented by OpenCL on Intel GPUs 
to formally motivate the need for CM.

\begin{enumerate}
\item \textit{Register file control}: 
Effective use of the register file to reduce unnecessary memory traffic is perhaps the most important optimization strategy for Intel GPUs \cite{chen2018register}. Careful management of register pressure is difficult to achieve in OpenCL, as its language leaves the decision of register allocation entirely in the compiler's hands. Hundreds of compiler transformation and optimization passes take place for an OpenCL kernel to be compiled into Gen assembly; most of them can have significant impact to register pressure, yet their behavior is nontransparent and usually non-controllable for the programmer. 

For example, divergence analysis \cite{coutinho2011divergence} is a critical analysis for SIMT GPU compilers, and its results may be used to reduce register usage by allocating a scalar register for a variable if can prove all lanes hold identical values. The analysis results are often overly conservative in the presence of complex data and control dependencies, but offers no mechanism for the programmer to assist the analysis. By contrast, CM variables are register-allocated by default, and vectors and matrices can have arbitrary size within hardware limit. CM developers can thus directly allocate their uniform variables in one register, and they may also coalesce variables into large matrices for explicit lifetime management.

\item \textit{Cross-lane data sharing}: 
A well-known limitation of the SIMT execution model is the lack of data sharing among the work-items in a hardware thread. Even though SIMD lanes in a thread share the register file, the SIMT abstraction prevents one lane from accessing another lane's register data, and this invariably leads to redundant computation and memory operations. Both CUDA and OpenCL have introduced explicit SIMD primitives to facilitate cross-lane communications, and functionalities provided include shuffle, reduction, and barrier operations \cite{cuda-warp, opencl-subgroup}. These extensions help bridge the gap between the SIMT model and the underlying SIMD hardware, but they do not represent actual hardware capabilities. By contrast, CM's select operation directly maps to hardware regioning and may be used directly in compute instructions, thus eliminating unnecessary shuffle moves. 


\item \textit{Vector length control}:
Each Gen ISA instruction has its own execution size, and per-instruction SIMD size can be an important optimization technique. One immediate use of varying vector size is register pressure control. Most applications go through phases of high and low register demand, and a kernel should mix its SIMD size to avoid spills in high-pressure regions while achieving maximum bandwidth for vector memory gather/scatter operations. Similarly, branch divergence can significantly reduce a program's efficiency\cite{10.1007/978-3-642-54807-9_8, han2011reducing}; in the absence of hardware mechanisms, the inactive channels will not execute until control flow re-converges. By running with a lower SIMD size inside divergent regions, a kernel could reduce the amount of wasted work.
Because of CM's explicit SIMD model, programmers can easily control each instruction's SIMD size through the size of vector and matrix selects. The SIMT model offers no such capabilities, however, as OpenCL GPU compilers perform implicit vectorization on the kernel. An OpenCL kernel may specify its dispatch size, but all non-uniform instructions will have that size by default.

\end{enumerate}

We use a simple 3 by 3 box blur filter (aka linear filter) to compare and contrast CM and OpenCL's programming models. We first show a 
straightforward OpenCL implementation and point out its efficiencies on Intel GPUs.
In Section \ref{section:programming-language} we present the CM implementation to showcase the language's key features, 
while Section \ref{section:compiler} explains how the CM kernel is compiled into the base ISA.
In Section \ref{section:experimental-evaluation}, we evaluate the performance of our CM kernel against an
optimized OpenCL kernel that uses Intel-specific extensions,
and show that even this optimized version can only reach less than 50\% of CM's performance. 

\begin{algorithm}
\caption{Linear filter in OpenCL with SIMT model}

\begin{algorithmic}[1]
\Procedure{linear}{image2d src, image2d dst, int width, int height} 
    \State int x = get\_global\_id(0);
    \State int y = get\_global\_id(1);
    \State float4 pixel1 = 0.0f;
    \State float4 pixel = 0.0f;
    \State int tempx, tempy;
    \Statex \textbf{$\#$pragma unroll}
	\For{$i=-1; i \leq 1; i$++}
		\Statex \textbf{$\#$pragma unroll}
		\For{$j = -1; j \leq 1; j$++}
			\State tempx = min(width-1, max(0, x+j));
			\State tempy = min(height-1, max(0, y+i));
			\State pixel1 = \textit{read}(src,sampler,(int2)(tempx,tempy));
			\State pixel.z += pixel1.z;
			\State pixel.y += pixel1.y;
			\State pixel.x += pixel1.x;
		\EndFor  
	
	\EndFor    
    \State uint4 p = convert\_uint4(pixel*0.1111f);
    \State \textit{write}(dst, (int2)(x,y), p);
\EndProcedure
	
\end{algorithmic}
\label{alg:linear-simt}
\end{algorithm}

In Algorithm \ref{alg:linear-simt}, every work-item computes the result of one pixel, whose position is indicated by 
the work-item's $x$ and $y$ global id, by taking the average value of its neighbors in the input image.
Intel's OpenCL compiler vectorizes this kernel into SIMD16 instructions where each lane corresponds to one pixel in the input and output image.
Both images are in 3-channel RGB format, and the hardware image read unit converts the 8-bit integer in each channel into 
normalized floating-point values in structure-of-array (SoA) format. 
The image write performs the format conversion in reverse. 
The generated assembly consists of 9 image-gather loads (line 11), 27 floating-point additions (line 12-14), and one image-scatter write (line 18).
 
This simple implementation suffers from severe redundant loads in each hardware thread, as in one iteration each work-item is reading pixel values that were 
already loaded in previous iterations by its adjacent lanes.  
A more efficient method is to have the work-items in a thread cooperatively load a 2D block of the image in raw format (i.e., the pixels are 
loaded into registers without format conversion), then convert each channel into floating-point values for subsequent computation.
This special 2D block read/write functionality is provided by Intel's cl\_intel\_media\_block\_io extension.
 
The effectiveness of this approach is still limited by the SIMT model, however, as the builtin function's return data must be 
evenly distributed among the work-items in a subgroup. Thus, a subgroup shuffle operation is required to read the neighbor lanes' 
pixels and convert them from array-of-structure (AoS) into SoA layout.
The OpenCL compiler is generally not able to optimize away these costly moves, as to satisfy the SIMT model it must maintain 
the values being computed in SoA format. As a last resort one could avoid the shuffle moves by transposing the input image
in host code, but this increases CPU overhead and real-world applications do not necessarily have control over their input layout.   

As we will show in the next section, these issues can be easily addressed in CM.
Since a CM kernel describes the algorithm for one thread, it can naturally store the data for the 2D block read/write in a matrix,
and it can also choose the best matrix size without being constrained by the dispatch size.
Explicit vectorization means CM developers can structure their code to accommodate the block load's layout,
and the select operations efficiently extract the sub-elements for computation.
The CM compiler's ability to break up matrix operations into variable-size Gen instructions simplifies programming efforts 
while maintaining high performance. 

\section{CM Programming Language}
\label{section:programming-language}

The CM programming language is implemented using Clang and supports a subset of the standard C++ with some restrictions (more details in section 2.6 of the CM language specification \cite{cm2018}). Two container types, \verb+vector+ and \verb+matrix+, are added to the Clang base type system. These new base types form the foundation for the CM explicit SIMD programming model. On top of these two types, we add operations and builtin functions that closely resemble the Gen instruction set. These new types and functions together form the abstract interface for close-to-the-metal programming on Gen. The following subsections illustrate the major features of the language. For all the details needed to write CM code, refer to the CM language specification \cite{cm2018}.

\subsection{Vector and Matrix Types}
These types are defined using syntax similar to C++ template classes. The parameters are the type of data element and the size of a vector/matrix. Element type must be one of the basic types supported by CM and sizes must be positive integers and compile-time constants.

\begin{Verbatim}[commandchars=\\\{\},fontsize=\small]
\textcolor{myblue}{vector}<short, 8> v;   \textcolor{mygreen}{// A vector of 8 shorts}
\textcolor{myblue}{matrix}<int, 4, 8> m;  \textcolor{mygreen}{// A 4x8 integer matrix}
\end{Verbatim}

Additionally, CM provides two reference component data types: \verb+vector_ref+ and \verb+matrix_ref+. They define references to basic vector or matrix objects. No extra memory space is allocated to reference variables. For example, the second row of matrix $m$ could be defined as a reference variable as:

\begin{Verbatim}[commandchars=\\\{\},fontsize=\small]
\textcolor{myblue}{vector_ref}<int, 8> vref(m.\textcolor{myblue}{row}(2));
\end{Verbatim}

Vector or matrix variables map to a sequence of consecutive elements residing in the general register file (GRF) of the Gen hardware.
A vector or matrix variable may not have its address taken; indirect access is performed via the reference types instead. 
Reference variables are usually constructed from operations on base variables which provide alternative views to the base objects.  Reading a reference variable is mapped directly to Gen’s region based addressing scheme, which provides zero-overhead data pack, unpack, and shuffling within two registers. 

For vectors, matrices, and their corresponding reference variables, CM supports member functions and operations including constructor and assignment; arithmetic, shift, logic and comparison; and row, column and element accesses.
The main operations unique to CM vector and matrix types are:
\begin{itemize}
\item \textit{\textbf{select:}} a set of select functions for referencing a subset of vector/matrix elements are supported. Each select operation returns a reference to the elements of the base object, and they can be used as \textit{l-value} expressions. Select operations are of the form (with \Verb[fontsize=\small]+v+ being a vector and \Verb[fontsize=\small]+m+ a matrix):

\begin{Verbatim}[commandchars=\\\{\},fontsize=\small]
v.\textcolor{myblue}{select}<size,stride>(i)
m.\textcolor{myblue}{select}<vsize,vstride,hsize,hstride>(i,j)
\end{Verbatim}

In the second case, it returns a reference to the sub-matrix starting from the (i, j)-th element. \textit{vsize} indicates the number of selected rows; \textit{vstride} indicates the distance between two adjacent selected rows; \textit{hsize} indicates the number of selected columns; and \textit{hstride} indicates the distance between two adjacent selected columns. As Figure \ref{fig:select} shows, \Verb[commandchars=\\\{\},fontsize=\small]+v.\textcolor{myblue}{select}<4, 2>(1)+ is an \textit{l-value} expression of type \Verb[commandchars=\\\{\},fontsize=\small]+\textcolor{myblue}{vector_ref}<float, 4>+, which refers to odd elements in the 8-float vector \Verb[fontsize=\small]+v+. In the case of matrix \Verb[fontsize=\small]+m+, the example shows that the  operation selects 4 elements (\textit{vsize}=2,  \textit{hsize}=2) with \textit{vstride} and \textit{hstride} of 2 and 4 respectively. The initial offset is \Verb[fontsize=\small]+m[1, 2]+.

\begin{figure}[ht!]
\centering
\vspace{2mm}
\includegraphics[width=\columnwidth]{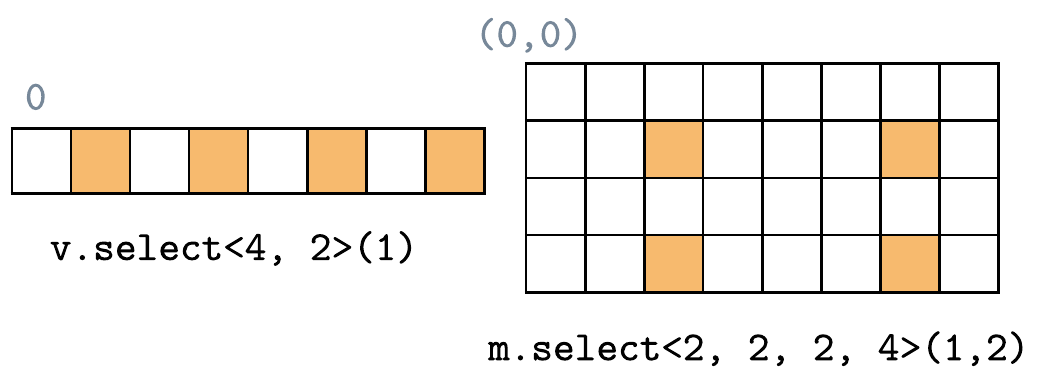}
\caption{Examples of select operation}
\label{fig:select}
\end{figure}

Nested vector or matrix select operations are efficiently mapped into direct register addressing operations on Gen.

\item \textbf{\textit{iselect:}} CM allows the user to perform indexed access into another vector. Indirect selects are always \textit{r-value} expressions. For example, consider a base variable \Verb[fontsize=\small]+v+ of 16 floats, and let \Verb[fontsize=\small]+idx+ be a vector of 4 elements $\{0, 1, 2, 2\}$. Then the expression \Verb[commandchars=\\\{\},fontsize=\small]+v.\textcolor{myblue}{iselect}(idx)+  can be used to create a new vector with elements $\{$\Verb[fontsize=\small]+v[0]+, \Verb[fontsize=\small]+v[1]+, \Verb[fontsize=\small]+v[2]+, \Verb[fontsize=\small]+v[2]+$\}$. This function exposes Gen's register-indirect addressing capability.

\item \textbf{\textit{merge:}} two forms of merge operations are provided to support conditional updates: \Verb[commandchars=\\\{\},fontsize=\small]+v.\textcolor{myblue}{merge}(x, mask)+ and \Verb[commandchars=\\\{\},fontsize=\small]+v.\textcolor{myblue}{merge}(x, y, mask)+. The former copies elements from \Verb[fontsize=\small]+x+ to \Verb[fontsize=\small]+v+ when the corresponding mask bit is true. The latter copies elements to \Verb[fontsize=\small]+v+ from \Verb[fontsize=\small]+x+ when the corresponding mask bit is true; otherwise, it copies elements to \Verb[fontsize=\small]+v+ from \Verb[fontsize=\small]+y+. The first merge is mapped to Gen's predicated \textit{mov} instructions, while the second merge is mapped to \textit{sel} instructions. 

\item \textbf{\textit{format:}} this operation allows reinterpreting the element type of a matrix/vector variable and changing its shape. 
As an example, on a vector \Verb[fontsize=\small]+v+ of 8 floats, the expression \Verb[commandchars=\\\{\},fontsize=\small]+v.\textcolor{myblue}{format}<char, 4, 8>()+ has type \linebreak \Verb[commandchars=\\\{\},fontsize=\small]+\textcolor{myblue}{matrix_ref}<char, 4, 8>+, meaning \Verb[fontsize=\small]+v+ is reinterpreted to a matrix of type char with 4 rows and 8 columns.

\item \textbf{\textit{replicate:}} this operation provides generic regioning operations to gather elements from a vector or matrix. The expression \Verb[commandchars=\\\{\},fontsize=\small]+v.\textcolor{myblue}{replicate}<K, VS, W, HS>(i)+ gathers \Verb[fontsize=\small]+K+ blocks from the input
vector \Verb[fontsize=\small]+v+ starting from position \Verb[fontsize=\small]+i+, and each block has \Verb[fontsize=\small]+W+ elements. \Verb[fontsize=\small]+VS+ and \Verb[fontsize=\small]+HS+ are the vertical and horizontal stride.  For example, \Verb[fontsize=\small]+v.replicate<2, 4, 4, 0>(2)+ on vector \Verb[fontsize=\small]+v+  from Figure \ref{fig:select} will gather the elements $\{$\Verb[fontsize=\small]+v[2]+, \Verb[fontsize=\small]+v[2]+, \Verb[fontsize=\small]+v[2]+, \Verb[fontsize=\small]+v[2]+, \Verb[fontsize=\small]+v[6]+, \Verb[fontsize=\small]+v[6]+, \Verb[fontsize=\small]+v[6]+, \Verb[fontsize=\small]+v[6]+$\}$.
\end{itemize}

CM also supports mixed operations of vector and matrix objects of different shapes as long as each operands has identical number of elements. The operand shape conformance is checked at compile time using template specialization rules for vector/matrix classes. The CM compiler determines the element type for the destination operand based on the source operand data types following standard C++ rules for type promotion (using template specialization mechanisms). 
Just like in standard C++, users may want to add explicit type conversions to change the default type promotion and conversion rules. 
A simple example of an implicit and explicit conversion can be:

\begin{Verbatim}[commandchars=\\\{\},fontsize=\small]
\textcolor{myblue}{vector}<float, 8> f;
\textcolor{myblue}{vector}<int, 8> i;
f = i;                   \textcolor{mygreen}{//Implicit conversion}
f = \textcolor{myblue}{vector}<short, 8>(i); \textcolor{mygreen}{//Explicit conversion}
\end{Verbatim}

CM allows vector and matrix to be declared as file-scope variables, which are treated as thread private variables. They can be used to facilitate data sharing among the main function and its callee functions in the same thread. Optionally, CM supports two variants of global variable usage. The first variant, denoted by the \Verb[commandchars=\\\{\},fontsize=\small]+_GENX_VOLATILE_+ qualifier, informs compiler to perform conservative optimizations on these variables in order to decrease register pressure and improve code quality. The second variant, denoted by the \Verb[commandchars=\\\{\},fontsize=\small]+_GENX_VOLATILE_BINDING_(Offset)+ qualifier, indicates the global variable should be mapped to a GRF block starting from the specified byte offset. Such register binding feature enables programmer to achieve fine-grained register allocation control and effectively tackle other challenges such as bank conflict for performance critical applications. 

\subsection{Memory Intrinsics}

CM provides a set of memory-access functions that resemble the underlying Gen hardware operations. 
By default a buffer-indexed based addressing mode is used. A kernel includes a number of \textit{SurfaceIndex} arguments, 
each of which represents a handle to the underlying memory object. A read or write intrinsic takes one surface index 
and accesses its elements specified by the offsets.
Application host code is responsible for binding each kernel argument to a memory object through runtime API calls.
The most useful intrinsics include:

\begin{itemize}

\item \textit{\textbf{2D-block read/write:}} For an image identified by its \textit{SurfaceIndex}, a block-read loads a block of pixels at the given x/y location into a matrix. A 2D-block write stores a matrix into a block of pixels in an image at the given x/y location. The following intrinsic definition is for \textit{2D-block read}.

\begin{Verbatim}[commandchars=\\\{\},fontsize=\small]
template<typename T, int N, int M>
void read(SurfaceIndex index,
          CmBufferAttrib attr, int X, int Y,
          \textcolor{myblue}{matrix_ref}<T, N, M> output)
\end{Verbatim}

\item \textit{\textbf{Oword-block read/write:}} For a linearly-addressed buffer, a block-read reads a consecutive sequence of owords (16 bytes per oword) 
at a given offset into a vector. A block-write writes a vector into a consecutive sequence of oword at the given offset into the buffer. The following intrinsic definition is for \textit{Oword-block read}.

\begin{Verbatim}[commandchars=\\\{\},fontsize=\small]
template<typename T, int N>
void read(SurfaceIndex idx,
          CmBufferAttrib attr, int offset, 
          \textcolor{myblue}{vector_ref}<T, N> output)
\end{Verbatim}

\item \textit{\textbf{Scattered read/write:}} Vector gather and scatter of various granularity are also supported. Zero-based offsets of each element (relative to a global offset) to be read/written are specified in a vector. For scattered read
and write functions, the address, source payload, and
return data must be vector type of the same size. The following intrinsic definition is for \textit{scattered read}.

\begin{Verbatim}[commandchars=\\\{\},fontsize=\small]
template <typename T, int N>
void read(SurfaceIndex index, 
          uint globalOffset, 
          \textcolor{myblue}{vector}<uint, N> elementOffset,
          \textcolor{myblue}{vector_ref}<T, N> ret) 
\end{Verbatim}

\item \textit{\textbf{Atomics:}}  CM supports all native atomic operations on Gen including \textit{and}, \textit{add}, \textit{max}, \textit{inc}, \textit{compxchg}, etc. Like scattered read/write, atomic functions 
must also have vector type. The following is the intrinsic definition for atomic \textit{inc}.

\begin{Verbatim}[commandchars=\\\{\},fontsize=\small]
template<CmAtomicOp Op, typename T, int N>
void write_atomic(\textcolor{myblue}{vector}<ushort, N> mask, 
            SurfaceIndex index, 
            \textcolor{myblue}{vector}<uint, N> element_offset)
\end{Verbatim}
\end{itemize}

In addition to \textit{SurfaceIndex}, CM also supports a flat addressing model where a kernel argument is a pointer that may be directly used for memory access. This allows 
host and kernel code to share data structures and concurrently access them. 

\subsection{Boolean Reductions}
To facilitate boolean reductions on mask vectors, CM provides two predefined boolean functions: 

\begin{Verbatim}[commandchars=\\\{\},fontsize=\small]
    ushort \textcolor{myblue}{vector}<ushort, size>::\textcolor{myblue}{any}(void)
    ushort \textcolor{myblue}{vector}<ushort, size>::\textcolor{myblue}{all}(void)
\end{Verbatim} 

\Verb[commandchars=\\\{\},fontsize=\small]+\textcolor{myblue}{any}()+ returns 1 if any of the value in the mask is non-zero; it returns 0 otherwise. \Verb[commandchars=\\\{\},fontsize=\small]+\textcolor{myblue}{all}()+ returns 1 if all the values in the mask are non-zero; it returns 0 otherwise. Notice that the same functions are also available for matrix types. The result of either function can be used as a scalar value and be used in the standard C++ control-flow constructs. Reduction functions are efficiently mapped to Gen's compare instructions.

\subsection{SIMD Control Flow}

In CM, the default control-flow statement is just the C++ scalar control flow statements – conditional statements (if-else/switch), loop statements (for/while/do-while), jump statements (break/continue/goto/return) or function calls. For those statements, the conditions must be scalars, and all SIMD lanes branch uniformly.

Beyond that, CM also provides per-lane SIMD control-flow mechanisms utilizing the Gen \verb+simd-goto+ and \verb+simd-join+ instructions that support divergent control-flow under SIMD execution \cite{chandrasekhar2019igc}. This feature provides an alternative to predicating long sequence of 
instructions, as inactive channels do not execute inside SIMD control flow regions. 

SIMD control flow in CM is expressed by predefined C++ macros. For instance, a divergent \textit{if} is represented by macros SIMD\_IF\_BEGIN and SIMD\_IF\_END, and are used as follows:

\begin{Verbatim}[commandchars=\\\{\},fontsize=\small]
  \textcolor{myblue}{vector}<uint, 16> v(0);
  \textcolor{myblue}{vector}<ushort, 8> cond = ...
  \textcolor{myblue}{SIMD_IF_BEGIN}(cond > 0)\{
    \textcolor{mygreen}{// ...}
    v.\textcolor{myblue}{select}<8, 2>(0) = 1;
  \}\textcolor{myblue}{SIMD_ELSE}\{
    \textcolor{mygreen}{// ...}
    v.\textcolor{myblue}{select}<8, 2>(1) = 1;
  \}\textcolor{myblue}{SIMD_IF_END};
\end{Verbatim}

The comparison $cond>0$ produces a vector mask that determines whether a lane is active. 
Both the \textit{then} statement and the \textit{else} statement may get executed for their active lanes. 
A SIMD control flow block is skipped if none of the lanes are active.  Notice that the size of SIMD operations within a SIMD control-flow must be either the same size as the mask or scalar.




\subsection{Linear Filter in CM}

\begin{figure}[h]
\centering
\includegraphics[width=\columnwidth]{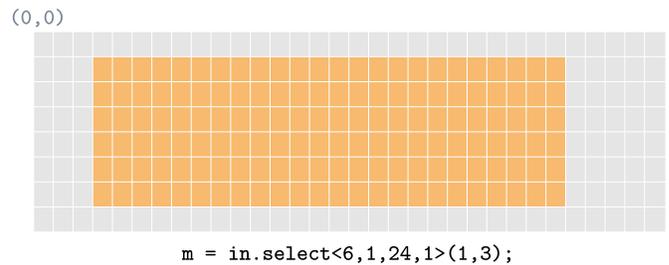}
\caption{Select a 6x24 sub-matrix from a 8x32 matrix}
\label{fig:linear-select}
\end{figure}

We now describe how the linear filter can be implemented in CM (Algorithm \ref{alg:linear-cm}).
Each thread in the CM kernel reads a 8x32-byte matrix and outputs a 6x24-byte matrix corresponding to 6x8 pixels. Although we only need 8x30 bytes for 8x10 input pixels,
adding two-byte padding to each row gives a good layout in register file for computation. 
The select operation acts as follows: after the input pixels are loaded into the 8x32-byte matrix \verb+m+, at each step, we extract a 6x24-byte sub-matrix through a select operation, convert all elements into float, then add them to the running total, which is a 6x24-floating matrix. Figure \ref{fig:linear-select} shows the first 6x24-byte sub-matrix select operation performed in Algorithm \ref{alg:linear-cm}.


\begin{algorithm}
\caption{Linear filter written in CM}
\begin{algorithmic}[1]
\Procedure{linear}{Surface inBuf, Surface outBuf, uint hpos, uint vpos}

    \State {\color{myblue}matrix}$<$uchar, 8, 32$>$ in;  \Comment{8x32 input matrix}
    \State {\color{myblue}matrix}$<$uchar, 6, 24$>$ out; \Comment{6x24 output matrix}
    \State {\color{myblue}matrix}$<$float, 6, 24$>$ m; 
    \State \textit{read}(inBuf, hpos*24, vpos*6, in); \label{alg:linear-cm:read}
    \State
    \Comment{Compute sums of neighbor elements}
    \State m = in.{\color{myblue}\textit{select}}$<$6, 1, 24, 1$>$(1, 3); \label{alg:linear-cm:sum1}
    \State m += in.{\color{myblue}\textit{select}}$<$6, 1, 24, 1$>$(0, 0);
    \State m += in.{\color{myblue}\textit{select}}$<$6, 1, 24, 1$>$(0, 3);
    \State m += in.{\color{myblue}\textit{select}}$<$6, 1, 24, 1$>$(0, 6);
    \State m += in.{\color{myblue}\textit{select}}$<$6, 1, 24, 1$>$(1, 0);
    \State m += in.{\color{myblue}\textit{select}}$<$6, 1, 24, 1$>$(1, 6);
    \State m += in.{\color{myblue}\textit{select}}$<$6, 1, 24, 1$>$(2, 0);
    \State m += in.{\color{myblue}\textit{select}}$<$6, 1, 24, 1$>$(2, 3);
    \State m += in.{\color{myblue}\textit{select}}$<$6, 1, 24, 1$>$(2, 6); \label{alg:linear-cm:sum2}
    \State
    \Comment{Compute average (implicit type conversion)}
    \State out = m*0.1111f; 
	\State \textit{write}(outBuf, hpos*24, vpos*6, out); \label{alg:linear-cm:write}
\EndProcedure
	
\end{algorithmic}
\label{alg:linear-cm}
\end{algorithm}

The 2D-block read/write functions are used to perform the load and store 
on line \ref{alg:linear-cm:read} and line \ref{alg:linear-cm:write}. 
As mentioned in Section \ref{section:motivations}, for this filter the specialized 2D block messages are 
much more efficient than the image gather/scatter operations in the vanilla OpenCL implementation (Algorithm \ref{alg:linear-simt})
due to the elimination of redundant memory traffic.

\section{CM Compiler}
\label{section:compiler}
Like Intel Graphics Compiler (IGC) \cite{chandrasekhar2019igc}, the CM Compiler consists of three layers:
 
\begin{itemize}
\item Front-end:  The clang front-end compiler \cite{lattner2008llvm} converts CM source code into LLVM intermediate representation (IR) \cite{lattner2004llvm}.
\item Middle-end: The middle-end performs generic and CM speciﬁc optimizations and transformations before converting the LLVM IR into the virtual-ISA (vISA) assembly language. The vISA is very close to Gen ISA but offers more convenience as a compilation target as it has unlimited virtual registers and hides various hardware-speciﬁc restrictions.
\item Finalizer: The vISA ﬁnalizer \cite{chen2018register} is a code generator for Intel GPU. Taking vISA assembly as input, it performs local optimizations, register allocation and scheduling to generate the ﬁnal instructions for the target Intel GPU.  
\end{itemize}

\begin{figure}[h]
\centering
\includegraphics[width=\columnwidth]{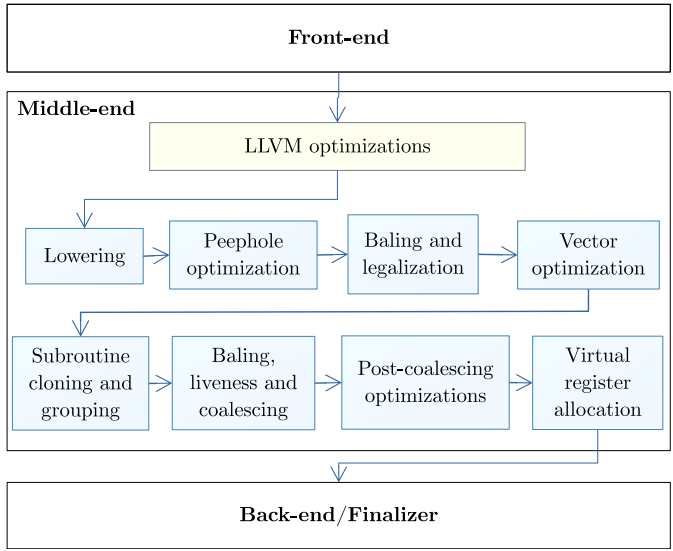}
\caption{CM compilation flow}
\label{fig:cmc}
\end{figure}

The general ﬂow of the CM custom optimizations is illustrated in Figure \ref{fig:cmc} (inside middle-end module). The input corresponds to LLVM IR generated by LLVM generic optimizations. The lowering pass gradually converts the high-level CM language constructs to code sequences that are closer to the target Gen ISA. Afterwards, several optimizations are performed at each IR level to improve the code quality. Two of these optimization passes are highlighted in the remainder of this section: bailing and legalization and vector optimization.

Gen ISA has distinct features such as varying execution size, mixed data types, flexible register regioning, and modifier support \cite{chandrasekhar2019igc}. Vector and matrix data types and their region-select operations need to be carefully modeled so that they can be directly mapped to those distinct features without extra move instructions. Since LLVM is based on Static Single Assignment (SSA) form, where each value is defined exactly once, we extend its IR with the following two intrinsics to model partial read/write to vector/matrix variables in SSA form, so that it can benefit from common LLVM optimizations.

\begin{itemize}
\item Read region (\textit{rdregion}): extract selected elements from a vector to make a new smaller vector.
\item Write region (\textit{wrregion}): insert elements into selected positions and returns a new value for the old vector.
\end{itemize}

The following is a simplified example to illustrate the design. The original vector \verb+a+ is defined as an \verb+8 x i32+ value \verb+%a0+. The \textit{rdregion} intrinsic extracts \verb+4 x i32+ elements from \verb+%a0+ based on the given parameters: vertical stride = 0, width = 4, horizontal stride = 2, starting byte offset = 4. The \textit{wrregion} intrinsic inserts the elements of \verb+%b+ to the old value of \verb+a+ (\verb+%a0+) based on the other given parameters: vertical stride = 0, width = 4, horizontal stride = 2, starting byte offset = 0. The SSA property is maintained as the \textit{wrregion} intrinsic returns a different \verb+%a1+ to represent the new value of vector \verb+a+.
 
\begin{Verbatim}[commandchars=\\\{\},fontsize=\small]
\textcolor{myblue}{vector}<int, 8> a(init_v);
\textcolor{myblue}{vector}<int, 4> b;
b = a.\textcolor{myblue}{select}<4, 2>(1);
a.\textcolor{myblue}{select}<4, 2>(0) = b;
 
%a0 = <8xi32> …
%b = call<4xi32> @llvm.genx.rdregioni...
 (<8xi32> %a0, i32 0, i32 4, i32 2, i16 4);
%a1 = call<8xi32> @llvm.genx.wrregioni...
 (<8xi32> %a0, <4xi32> %b, i32 0,
 i32 4, i32 2, i16 0);
\end{Verbatim}

Due to its expressiveness one vISA instruction may be represented in the LLVM IR by multiple instructions. Baling is the process of determining which group of LLVM instructions can be combined (baled) together and efficiently mapped to vISA. A bale has a root instruction as well as optional modifiers and region instructions on the source and destination operands. The baling analysis pass constructs a map to mark which IR instructions are selected and what roles they play in their resulting bales. The root of a bale is the last instruction in the program order of all instructions in the bale, which is also the only instruction whose value is used outside the bale. Since the baling pass may decide to bale in an instruction with multiple uses as a non-root instruction, the instruction is cloned to ensure it has only a single use inside the bale.
 
vISA is designed to be close to Gen ISA and inherits similar restrictions (e.g., the size of an operand may not exceed two GRFs). 
After the initial baling analysis, the legalization pass may split up one bale into multiple instructions to conform to vISA restrictions. 
In general, the splitting must be done carefully to take advantage of the maximum SIMD width allowed by the target platform. 
Other examples of transformations performed here include un-baling an instruction due to conflicting legalization requirements, 
aligning operands for memory access operations, and promoting byte type operations into equivalent short ones to work around hardware restrictions.
 
The vector optimization pass performs optimizations based on \textit{rdregion} and \textit{wrregion} tailored for vector and matrix. The following are a few examples:

\begin{itemize}
\item Constant folding: We have extended LLVM constant folding so that it can fold and propagate vector constants through \textit{rdregions} and \textit{wrregions}.
\item Promoting C-array into LLVM vector: Although it is not recommended, users can use a C-array in CM instead of a CM vector. The CM compiler can replace C-array loads and stores with \textit{rdregions} and \textit{wrregions}.
\item Region collapsing: This can be viewed as instruction-combining transformation specific to \textit{rdregions} and \textit{wrregions}.
\item Dead vector removal: This is a more general form of dead-code elimination on vector values. The uses of every vector element are tracked to determine if the whole vector is dead.  
\item Vector decomposition: Given a large vector, if compiler can show that it can be divided into multiple segments, where the \textit{rdregions} and \textit{wrregions} on these segments are disjoint, then this large vector can be converted into multiple small ones, which increases the flexibility for the register allocator.
\end{itemize}

\begin{figure}[h]
\centering
\includegraphics[width=\columnwidth]{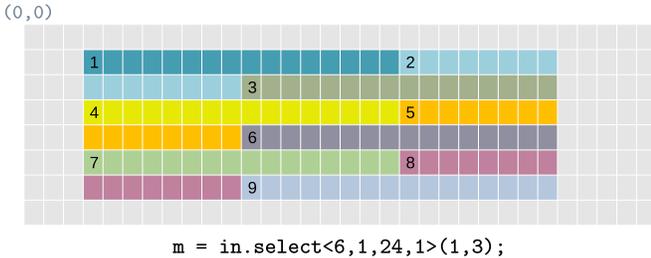}
\caption{Sub-matrix layout of a 6x24 char-to-float select operation.}
\label{fig:linear-region}
\end{figure}

As an example of the compiler code generation, consider again the linear CM implementation presented in Algorithm \ref{alg:linear-cm}. Figure \ref{fig:linear-region} illustrates how a 6x24 sub-matrix char-to-float conversion is done through a select operation (line \ref{alg:linear-cm:sum1} in Algorithm \ref{alg:linear-cm}).

This select operation is compiled into 9 SIMD16 instructions as shown below: 

\begin{Verbatim}[fontsize=\small]
1) mov (16|M0) r11.0<1>:f r4.3<8;8,1>:ub
2) mov (16|M0) r13.0<1>:f r4.19<16;8,1>:ub
3) mov (16|M0) r15.0<1>:f r5.11<8;8,1>:ub
4) mov (16|M0) r17.0<1>:f r6.3<8;8,1>:ub
5) mov (16|M0) r19.0<1>:f r6.19<16;8,1>:ub
6) mov (16|M0) r21.0<1>:f r7.11<8;8,1>:ub
7) mov (16|M0) r23.0<1>:f r8.3<8;8,1>:ub
8) mov (16|M0) r25.0<1>:f r8.19<16;8,1>:ub
9) mov (16|M0) r27.0<1>:f r9.11<8;8,1>:ub
\end{Verbatim}

In Gen ISA, a source operand's region is a
2D-array in row-major order with the format $<$V;W,H$>$, where
W (width) is the number of elements in a row, H (horizontal
stride) is the step size between two elements in a row, and V
(vertical stride) is the step size between two rows.
This example shows the power of CM programming on Gen; programmers 
express their algorithms using high-level matrix operations, and
the compiler generates them into multiple SIMD instructions while taking advantage of the 
region-based address scheme to efficiently access register data.

\section{Experimental Evaluation}
\label{section:experimental-evaluation}

This section presents a set of applications from different domains implemented in CM and OpenCL with their experimental evaluation on an Intel GPU. We also analyze results in terms of the productivity and development effort from the development process of several compute kernels.

\subsection{Applications}

We briefly highlight the implementation strategy of every CM kernel that enables them to achieve close-to-the-metal performance. The source code and description of the applications benchmarked can be found in \cite{cm2018} and in the appendix of this paper. The OpenCL kernels are from the Intel OpenCL SDK \cite{opencl-sdk} except for histogram and k-means which were 
developed internally by expert OpenCL programmers. All of them have been tuned and represent state-of-the-art 
OpenCL implementations for Intel GPUs. As baseline, all kernels were compiled with -O2 for the optimization level.

Typical input parameters were used for benchmarking the applications and their specification is described in every subsection; a detailed study of application behavior with varying input sizes is beyond the scope of this paper.

\begin{figure*}[!h]
\centering
\includegraphics[scale=1.15]{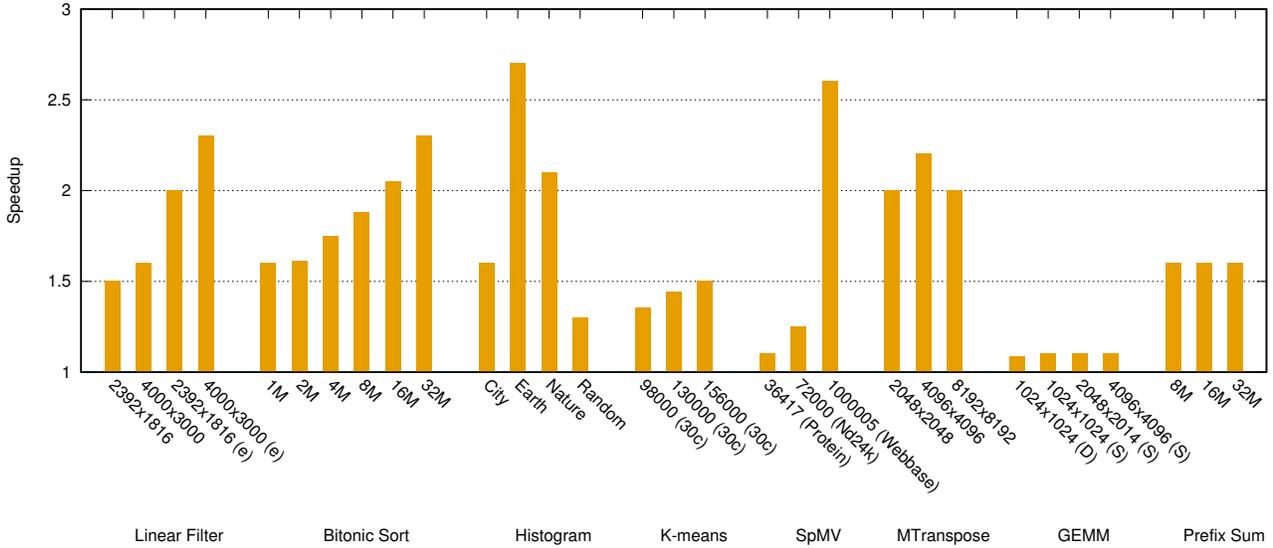}
\caption{Speedup of CM versus OpenCL kernels. Speedup is computed as $\frac{OpenCL\_exec\_time}{CM\_exec\_time}$.}
\label{fig:speedup}
\end{figure*}

The Intel IceLake (ICL) processor was used to run the workloads. The ICL system includes an Intel Core i7 with 4 CPU cores, 16GB of system memory and a Gen11 integrated GPU with 64 EUs. Performance comparison is done by measuring the total execution time.

\begin{enumerate}
\item \textbf{Bitonic Sort:} it is a classic parallel algorithm for sorting elements \cite{lee1996bitonic}. Given $2^n$ input elements, the bitonic network takes $n$ stages to sort, producing chunks of sorted elements in ascending and descending order in every stage. At every stage there is a split procedure that cuts one bitonic sequence into two smaller ones. The SIMT bitonic sort implementation benefits from using vector data types (e.g. int4) available in OpenCL, however, it involves global memory access within every stage. To avoid excessive global memory access and global synchronizations, our CM kernel takes advantage of the large register space to hold 256 data elements in registers, processing several split steps locally. Experimental results show that our CM implementation outperforms the OpenCL version by 1.6x to 2.3x as shown in Figure \ref{fig:speedup}.  The higher speedup with larger input sizes is due to additional savings from memory accesses and global synchronizations. 

\item \textbf{Histogram:} it is a common statistical tool used in image processing applications. It
collects the distribution of pixel intensities from an image. Both CM and OpenCL are based on local and global histograms to perform the parallel computation. However, while in the OpenCL implementation each thread's local histogram is stored in the SLM, in the CM kernel it is efficiently stored in registers. Also, in the OpenCL kernel one additional step is needed: after the local histogram computation the first thread in a work-group atomically updates the global histogram with local results. Figure \ref{fig:speedup} shows that CM significantly outperforms OpenCL, achieving up to 2.7x speedup. Furthermore, OpenCL's performance is very sensitive to different input patterns. The performance gap is narrower for randomly-generated input, where the OpenCL kernel is unlikely to incur SLM bank conflicts and serialized atomic increments. For real-world images with homogeneous background (e.g., earth), however, OpenCL's performance degrades significantly due to contention among atomic operations.

\item \textbf{K-means Clustering:} it is a popular clustering algorithm used in data mining and machine learning \cite{alsabti1997efficient}. K-means stores $k$ centroids that it uses to define clusters. A point is considered to be in a particular cluster if it is closer to that cluster's centroid than any other centroid. 
The CM k-means kernel is divided into two phases that iterate alternatively until the centroids converge. 
The first phase
divides input data into chunks of elements. Each hardware
thread processes the clustering for each chunk and computes
the minimum distance to determine which cluster (centroid) a
point belongs. The second phase sums up the accumulated coordinates and the number of points in each cluster and computes the new
centroid positions. In a final step, coordinates of the thread's cluster are produced. Compared to the OpenCL implementation, in Figure \ref{fig:speedup} it can be seen that the CM k-means is 30\% to 50\% faster with three different data sets. This performance difference is mainly because the CM k-means efficiently shares centroids and other auxiliary data structures in the register file instead of using SLM and thread barriers. The CM kernel also benefits from efficient scattered memory reads, which are overlapped by the CM compiler for latency hiding. 

\item \textbf{Sparse Matrix-Vector Multiplication (SpMV):}   for a sparse matrix $A$, SpMV computes the result of $Y = AX$, where $Y$ and $X$ are two dense vectors. It is widely used in many graph algorithms and scientific applications. The SIMT OpenCL implementation uses the cl\_intel\_subgroup extension and SLM efficiently, however, the presence of irregular memory accesses due to the nature of the input limits its performance. The CM implementation tackles this issue by adding the capability of dynamically varying the instruction SIMD. Since  issuing wider vector loads than
necessary wastes memory bandwidth and increases contention, we use dynamic branches to check different block sizes and select the best execution size accordingly. This capability
of varying SIMD size to improve both
memory and compute efficiency is an important CM advantage
over OpenCL. Another advantage is the use of boolean reductions that are applied to detect if all input rows are
zero and skip the entire computation. This also improves both memory and compute efficiency for sparse matrices. Experimental results in Figure \ref{fig:speedup} show that the CM kernel outperforms the OpenCL implementation by 10\% and  25\% for the Protein and Nd24k matrices which have the highest number of non-zero elements per row (around 200). For Webbase which has low density and high variance of non-zero elements (3 non-zeros/row), varying SIMD width is effective on achieving high memory efficiency and it performs 160\% better than OpenCL.

\item \textbf{Matrix Transpose:} it is a fundamental linear algebra operation that is heavily used in machine learning workloads. An optimized SIMT GPU implementation \cite{harris2013efficient} typically utilizes the SLM to avoid uncoalesced global memory access. For an out-of-place matrix transpose, threads within a thread group cooperatively copy a tile of the matrix from global memory into SLM, perform barrier synchronization, then copy SLM data using transposed array indices to the global output buffer. The CM implementation  can completely bypass SLM and avoid synchronization overhead by directly performing the transpose on registers. Transpose is performed using a combination of CM’s select and merge operations to shuffle each element to their transposed position. For example, the following CM code sequence transposes a $2 \times 2$ matrix
$m = \begin{bmatrix}
    a       & b \\
    c       & d
\end{bmatrix}$:
\begin{Verbatim}[commandchars=\\\{\},fontsize=\small]
v0 = v.\textcolor{myblue}{replicate}<2,1,2,0>(0); \textcolor{mygreen}{// [a,a,b,b]}
v1 = v.\textcolor{myblue}{replicate}<2,1,2,0>(2); \textcolor{mygreen}{// [c,c,d,d]}
v2 = \textcolor{myblue}{merge}(v0, v1, 0b0101);   \textcolor{mygreen}{// [a,c,b,d]}
\end{Verbatim}

We view $m$ as a vector $v = [a, b, c, d]$ and $v_2$ as the transpose of the original input matrix. Transpose of bigger matrices can be solved by 
recursively applying the above steps to each sub-matrix.

Experimental results on different matrix sizes, as illustrated in Figure \ref{fig:speedup}, 
show that this CM implementation achieves a speedup of up to 2.2x compared to the SLM-based OpenCL implementation.
OpenCL's subgroup shuffle functions do not help here since they are not expressive enough to exploit Gen's operand regioning.

\item \textbf{SGEMM and DGEMM:} General Matrix-to-Matrix Multiplication (GEMM) is a function that performs matrix multiplication of the form $C = \alpha A B + \beta C$,
where $A$, $B$ and $C$ are dense matrices and $\alpha$ and $\beta$ are scalar coefficients.
It is at the heart of many scientific applications and achieving
peak theoretical performance is critical for every architecture. Here we focus on single precision floating-point
(SGEMM) and double precision floating-point (DGEMM). Even though OpenCL and CM GEMM kernels employ a similar register-blocking strategy --OpenCL is able to do so by using the cl\_intel\_subgroup extension \cite{sgemm2015ocl} and mimicking the CM implementation, the CM kernel is able to process more data per thread thanks to more efficient management of the register file. As a result, CM outperforms OpenCL by 8.5\% in DGEMM and around 10\% in SGEMM for different input sizes as illustrated in Figure \ref{fig:speedup}. 

\item \textbf{Prefix Sum:} it is the cumulative sum of a sequence of numbers and plays an important role in many algorithms, e.g., stream compaction, radix sort, etc. The OpenCL implementation is based on Blelloch's algorithm \cite{blelloch1989scans} and uses a tree-traversal approach to build the prefix sum with parallel reductions and partial sums. It exploits the SLM but incurs several data movements between local and global memory, plus multiple barriers. Our CM implementation uses a similar approach but threads perform the parallel reduction and partial sums entirely in registers, updating their results  in place on the input array through scattered writes.  Figure \ref{fig:speedup} depicts that the CM implementation achieves 1.6x speedup compared to the OpenCL kernel for different input sizes.
\end{enumerate}

\subsection{Productivity}
Programmability is a common concern for the adoption of close-to-the-metal programming models, as one must carefully weigh their performance advantages against the potential developer productivity loss due to the ramp-up overhead and a lower level of abstraction. CM has been extensively used for high-performance library development inside Intel, however, and user experiences overwhelmingly suggest that programmers are much more productive using CM once performance tuning efforts are considered. During the early stages of kernel development for Intel’s deep learning neural network libraries, there was an intense debate on the choice of programming model. To ensure a fair comparison, a team of GPU compute architects implemented several key kernels in both OpenCL and CM. The architects in the study have years of experiences developing workloads in both models for Intel GPUs. Table \ref{table:productivity} details the development efforts as well as the performance achieved by both programming models. Development effort is measured as the amount of work performed to implement each kernel from scratch and meet the minimal performance requirement. Performance data are collected on a simulator for a future GPU platform and thus not included in the evaluation earlier in this section. Performance speedup is calculated as $\frac{OpenCL\_exec\_time}{CM\_exec\_time}$.

\begin{table}[!h]
\caption{Development effort and performance comparison.}
\begin{tabular}{ | C{1.8cm} C{1.85cm} C{1.85cm} C{1.65cm} | }
\hline
 Kernel & OCL effort (person-week) & 
 CM effort (person-week) & Performance (OCL/CM) \\ \hline \hline
 Systolic GEMM & 8 & 3 &  1.09x \\ \hline
 DGEMM and SGEMM & 12 & 4 &  1.06$\sim$1.09x \\ \hline
 Conv. 1x1 & 4 & 4 & 1.08x  \\ \hline
 Conv. 3x3 & 15 &4 & 1.3x  \\ \hline
 Stencil2D & 2$\sim$3 & 1 & 2.2x  \\ \hline
\end{tabular}
\vspace{0.2cm}
\label{table:productivity}
\end{table}

Table \ref{table:productivity} shows that for these deep learning kernels CM yields 2-3x more productivity than OpenCL on average while achieving better performance.The study found that developers could deliver functional OpenCL kernels quickly, but the initial version’s performance is often far below the desired targets. During the subsequent performance tuning, they have to spend considerable efforts fighting with the programming model and the compiler to get the desired assembly code. To achieve the best performance, developers need to control multiple aspects of kernel behavior including register usage, data sharing, latency hiding, copy coalescing, and bank conflict avoidance. The SIMT abstraction makes it difficult for even expert GPU programmers to control a kernel’s full optimization needs, and their OpenCL implementation suffers from poor performance predictability; an innocuous one-line change could result in significant variation in generated code if it causes the kernel to spill or copy moves to not be coalesced. On the contrary, CM allows users to manage critical machine resource explicitly to instruct the compiler to generate expected code sequence. The first working CM version is frequently able to approach or sometimes even exceed the performance target, thus greatly reducing the need for intensive tuning and rewrites later.

\section{Conclusions}
\label{section:conclusions}
This paper presents C-for-Metal, a high-level yet close-to-the-metal programming language for Intel GPUs. Major features are illustrated for how to expose underlying hardware capabilities: {\it vector/matrix} variables represent registers and express SIMD parallelism, {\it select} operation maps to register regioning, {\it block read/write} enables efficient memory access, and divergent control flow constructs allow for mixing SIMT and SIMD models. We evaluate several applications and their experimental results show that the performance gap between CM and OpenCL can be significant, ranging from 20\% to over 100\%. 
 
This paper is not meant to be an attack on SIMT programming models; they are popular on GPUs for a reason and several of the authors are active contributors to Intel's OpenCL compiler. 
Rather, we have shown that the convenience of the SIMT abstraction carries a performance cost that can be difficult to overcome even with expert programming.
A programming model that is natively designed to harvest hardware capabilities fully thus fills an essential void, 
and this metal-level expressiveness is especially important for performance-critical applications.


CM is positioned as a low-level programming tool for Intel GPUs. Different languages' front ends have started using CM as their back end. For instance, DPC++-ESIMD \cite{dpcpp-esimd} integrates some CM language features into DPC++, and ISPC \cite{ispc-gen} also generates CM vector intrinsics and relies on CM optimizations and code generation. Moreover, given the rising importance of vector and matrix data types for neural-network programming, we foresee that IR extensions similar to our \textit{rdregion} and \textit{wrregion} may be added into LLVM for other target machines.

\section*{Acknowledgment}
We thank many colleagues
who supported the CM compiler project and contributed to its development over the past years, including  Tim Corringham, Zhenying Liu, Wei Pan, Tim Renouf, David Stuttard, and Stephen Thomas. We also thank the anonymous
reviewers for their suggestions and comments.

\appendix
\section{Artifact Appendix}

\subsection{Abstract}

Our artifact contains the implementation of the CM compiler (CMC) as well as the applications and benchmarks used in the experimental evaluation section. We provide the required scripts to compile and execute the benchmarks, which allows the reproducibility  of our results on any system with Intel Gen9 (Skylake) GPU or above. 

\subsection{Artifact Meta-Information}

{\small
\begin{itemize}
  \item {\bf Program:} The CM compiler implemented in C++; CM applications; OpenCL applications (all sources and binaries included).
  \item {\bf Compilation:} With provided scripts via gcc/g++.
  \item {\bf Data set:} Applications use input data sets included either as separated files or generated at runtime. For the former case, they are located in each application directory.
  \item {\bf Run-time environment:} Linux Ubuntu 18.04 or above, CM runtime and OpenCL runtime.
  \item {\bf Hardware:} Intel Gen9 GPU or above.
  \item {\bf Output:} Performance results in text files for every application evaluated with CM and OpenCL.
  \item {\bf Publicly available:} The CM compiler as well as all the CM and OpenCL examples are publicly available except from those listed in the productivity section (section 6.1).
  \item {\bf Code license: } The Intel(R) CM compiler and examples are distributed under the MIT license.
\end{itemize}

\subsection{Description}

\subsubsection{How Delivered}

The CM compiler is available on Github: \url{https://github.com/intel/cm-compiler}. The CM and OpenCL examples, as well as scripts to build and run all the benchmarks are available on \url{https://github.com/jfuentes/C-for-Metal_CGO2021}. Binaries of the CM compiler and benchmarks are also included in the artifact repository.

\subsubsection{Hardware Dependencies}
We recommend running the benchmarks on an Intel Gen11 GPU (Icelake), however, any other Intel GPU above Gen9 (Skylake) should give similar results. Notice that due to hardware configuration differences, further application-specific tuning may be required to achieve peak performance on different Gen platforms.

\subsubsection{Software Dependencies}
This artifact was prepared using Ubuntu 18.04. Similar Linux distributions should also work. The artifact repository contains the CM compiler build and its dependencies to compile all the benchmarks. To build the CM and IGC compilers from sources, specific details about dependencies and how to build them can be found in their repositories: 
\begin{itemize}
\item CMC: \url{https://github.com/intel/cm-compiler}
\item IGC: \url{https://github.com/intel/intel-graphics-compiler}
\end{itemize}

To run the benchmarks the CM runtime and OpenCL runtime are required, which can be found in their repositories:
\begin{itemize}
\item CM runtime: \url{https://github.com/intel/media-driver}
\item OpenCL oneAPI Level Zero Runtime: \url{https://github.com/intel/compute-runtime}
\end{itemize}


\subsection{Installation}


First, install elemental dependencies for this artifact: g++, git, make, cmake and jansson.
\begin{Verbatim}[commandchars=\\\{\},fontsize=\small]
$ sudo apt install g++ git git-lfs make cmake
  libjansson-dev
\end{Verbatim}

\subsubsection{CM Compiler, Runtime and Benchmarks}
Download the artifact repository. It contains a build of the CM compiler and all the benchmarks. If building the CM compiler from sources is preferred, visit the CM compiler repository for more details (\url{https://github.com/intel/cm-compiler}). Also, notice that some applications files are uploaded via lfs. So make sure they are downloaded properly.
\begin{Verbatim}[commandchars=\\\{\},fontsize=\small]
$ git clone
https://github.com/jfuentes/C-for-Metal_CGO2021
$ cd C-for-Metal_CGO2021
$ git lfs pull
\end{Verbatim}

\noindent Now, we need to build and install the media driver which contains the CM runtime needed to run CM applications. Install prerequisites:
\begin{verbatim}
$ sudo apt install autoconf libtool libdrm-dev 
  xorg-dev openbox libx11-dev libgl1-mesa-glx 
  libgl1-mesa-dev xutils-dev
\end{verbatim}

\noindent Build and install libva:
\begin{verbatim}
$ git clone https://github.com/intel/libva.git
$ cd libva
$ ./autogen.sh --prefix=/usr 
  --libdir=/usr/lib/x86_64-linux-gnu
$ make
$ sudo make install
\end{verbatim}

\noindent Finally, build the media driver:
\begin{verbatim}
$ git clone 
  https://github.com/intel/media-driver.git
$ git clone https://github.com/intel/gmmlib.git
$ mkdir build_media & cd build_media
$ cmake ../media-driver/
$ make -j8
$ sudo make install 
\end{verbatim}

\noindent Notice that at this point you might need to set the path of the driver and make sure the path for dynamic libraries is set:
\begin{verbatim}
$ export LIBVA_DRIVERS_PATH=/usr/lib/
  x86_64-linux-gnu/dri
$ export LIBVA_DRIVER_NAME=iHD
$ LD_LIBRARY_PATH=$LD_LIBRARY_PATH:/usr/
  local/lib
$ export LD_LIBRARY_PATH
\end{verbatim}

\subsubsection{OpenCL Compiler (IGC) and Runtime for Intel GPU}
To install IGC and NEO runtime download the packages and follow the instructions from the compute runtime repository at \url{https://github.com/intel/compute-runtime/releases}.

\noindent Then, install OpenCL headers:
\begin{verbatim}
$ git clone https://github.com/KhronosGroup/
  OpenCL-Headers.git
$ cd OpenCL-Headers
$ sudo mv CL/ /usr/include/
\end{verbatim}

\noindent Additionally, you need to install the OpenCL C++ headers. Follow the installation steps from \url{https://github.com/KhronosGroup/OpenCL-CLHPP}.

\noindent Finally, install the OpenCL Installable Client Driver (ICD)
\begin{verbatim}
$ git clone https://github.com/KhronosGroup/
  OpenCL-ICD-Loader.git
$ cd OpenCL-ICD-Loader 
$ mkdir build & cd build 
$ cmake ..
$ make 
$ sudo make install
\end{verbatim}

\subsection{Experiment Workflow}
Once the above packages are installed, all the CM and OCL benchmarks can be built. Locate at the artifact repository and simply run:
\begin{verbatim}
$ cd benchmarks
$ sh build_CM_all.sh
$ sh build_OCL_all.sh
\end{verbatim}

\noindent The above command will generate both the kernel binaries and host executables for every benchmark. Notice that as the CM compilation  is offline compilation it will ask the GPU platform you are compiling for (SKL, ICL, etc.). Then, run the benchmarks:
\begin{verbatim}
$ sh run_CM_all.sh
$ sh run_OCL_all.sh
\end{verbatim}


\subsection{Evaluation and Expected Result}
Once the benchmarks are finished, performance results are reported to the standard output as well as text files located in the {\em results} directory. For each benchmark the kernel execution time and total execution time are reported. Performance results are in milliseconds and organized by input data.






\bibliographystyle{IEEEtran}
\bibliography{IEEEabrv,references}

\begin{thebibliography}{10}
\providecommand{\url}[1]{#1}
\csname url@samestyle\endcsname
\providecommand{\newblock}{\relax}
\providecommand{\bibinfo}[2]{#2}
\providecommand{\BIBentrySTDinterwordspacing}{\spaceskip=0pt\relax}
\providecommand{\BIBentryALTinterwordstretchfactor}{4}
\providecommand{\BIBentryALTinterwordspacing}{\spaceskip=\fontdimen2\font plus
\BIBentryALTinterwordstretchfactor\fontdimen3\font minus
  \fontdimen4\font\relax}
\providecommand{\BIBforeignlanguage}[2]{{%
\expandafter\ifx\csname l@#1\endcsname\relax
\typeout{** WARNING: IEEEtran.bst: No hyphenation pattern has been}%
\typeout{** loaded for the language `#1'. Using the pattern for}%
\typeout{** the default language instead.}%
\else
\language=\csname l@#1\endcsname
\fi
#2}}
\providecommand{\BIBdecl}{\relax}
\BIBdecl

\bibitem{nickolls2008scalable}
J.~Nickolls, I.~Buck, M.~Garland, and K.~Skadron, ``Scalable parallel
  programming with {CUDA},'' \emph{Queue}, vol.~6, no.~2, pp. 40--53, 2008.

\bibitem{munshi2009opencl}
A.~Munshi, ``The {OpenCL} specification,'' in \emph{2009 IEEE Hot Chips 21
  Symposium (HCS)}.\hskip 1em plus 0.5em minus 0.4em\relax IEEE, 2009, pp.
  1--314.

\bibitem{lattner2004llvm}
C.~Lattner and V.~Adve, ``{LLVM}: A compilation framework for lifelong program
  analysis \& transformation,'' in \emph{International Symposium on Code
  Generation and Optimization, 2004. CGO 2004.}\hskip 1em plus 0.5em minus
  0.4em\relax IEEE, 2004, pp. 75--86.

\bibitem{neo2020}
{Intel Corporation}, ``{Intel(R) Graphics Compute Runtime for oneAPI Level Zero
  and OpenCL(TM) Driver},'' \url{https://github.com/intel/compute-runtime},
  2020.

\bibitem{oneapi}
\BIBentryALTinterwordspacing
------, \emph{{oneAPI Level Zero Specification}}, 2020. [Online]. Available:
  \url{https://spec.oneapi.com/level-zero/latest/index.html}
\BIBentrySTDinterwordspacing

\bibitem{cm2018}
------, ``{C-for-Metal Compiler},'' \url{https://github.com/intel/cm-compiler},
  2019.

\bibitem{subgroup}
\BIBentryALTinterwordspacing
------, \emph{{Intel Subgroup Extension Specification}}, 2016. [Online].
  Available:
  \url{https://www.khronos.org/registry/OpenCL/extensions/intel/cl_intel_subgroups.html}
\BIBentrySTDinterwordspacing

\bibitem{1016024}
\BIBentryALTinterwordspacing
S.~Rul, H.~Vandierendonck, J.~D'Haene, and K.~De~Bosschere,
  ``\BIBforeignlanguage{eng}{An experimental study on performance portability
  of {OpenCL} kernels},'' in \emph{\BIBforeignlanguage{eng}{Application
  Accelerators in High Performance Computing, 2010 Symposium, Papers}}, 2010,
  p.~3. [Online]. Available:
  \url{http://saahpc.ncsa.illinois.edu/papers/paper_2.pdf}
\BIBentrySTDinterwordspacing

\bibitem{du2012cuda}
P.~Du, R.~Weber, P.~Luszczek, S.~Tomov, G.~Peterson, and J.~Dongarra, ``{From
  CUDA to OpenCL: Towards a performance-portable solution for multi-platform
  GPU programming},'' \emph{Parallel Computing}, vol.~38, no.~8, pp. 391--407,
  2012.

\bibitem{pennycook2013investigation}
S.~J. Pennycook, S.~D. Hammond, S.~A. Wright, J.~Herdman, I.~Miller, and S.~A.
  Jarvis, ``{An investigation of the performance portability of OpenCL},''
  \emph{Journal of Parallel and Distributed Computing}, vol.~73, no.~11, pp.
  1439--1450, 2013.

\bibitem{10.1007/978-3-642-38750-0_11}
Y.~Zhang, M.~Sinclair, and A.~A. Chien, ``Improving performance portability in
  opencl programs,'' in \emph{Supercomputing}, J.~M. Kunkel, T.~Ludwig, and
  H.~W. Meuer, Eds.\hskip 1em plus 0.5em minus 0.4em\relax Berlin, Heidelberg:
  Springer Berlin Heidelberg, 2013, pp. 136--150.

\bibitem{7284453}
T.~L. {Falch} and A.~C. {Elster}, ``Machine learning based auto-tuning for
  enhanced opencl performance portability,'' in \emph{2015 IEEE International
  Parallel and Distributed Processing Symposium Workshop}, 2015, pp.
  1231--1240.

\bibitem{10.5555/2066302}
\BIBentryALTinterwordspacing
J.~Fang, A.~L. Varbanescu, and H.~Sips, ``A comprehensive performance
  comparison of cuda and opencl,'' in \emph{Proceedings of the 2011
  International Conference on Parallel Processing}, ser. ICPP '11.\hskip 1em
  plus 0.5em minus 0.4em\relax USA: IEEE Computer Society, 2011, p. 216–225.
  [Online]. Available: \url{https://doi.org/10.1109/ICPP.2011.45}
\BIBentrySTDinterwordspacing

\bibitem{intel-intrinsics}
\BIBentryALTinterwordspacing
{Intel Corporation}, \emph{{Intel Intrinsics Guide}}, 2020. [Online].
  Available:
  \url{https://software.intel.com/sites/landingpage/IntrinsicsGuide/}
\BIBentrySTDinterwordspacing

\bibitem{c++-simd}
\BIBentryALTinterwordspacing
{C++ Standards Committee}, \emph{{Data-parallel vector library}}, 2020.
  [Online]. Available:
  \url{https://en.cppreference.com/w/cpp/experimental/simd}
\BIBentrySTDinterwordspacing

\bibitem{10.1145/2870650.2870653}
\BIBentryALTinterwordspacing
A.~Pohl, B.~Cosenza, M.~A. Mesa, C.~C. Chi, and B.~Juurlink, ``{An Evaluation
  of Current SIMD Programming Models for C++},'' in \emph{Proceedings of the
  3rd Workshop on Programming Models for SIMD/Vector Processing}, ser. WPMVP
  ’16.\hskip 1em plus 0.5em minus 0.4em\relax New York, NY, USA: Association
  for Computing Machinery, 2016. [Online]. Available:
  \url{https://doi.org/10.1145/2870650.2870653}
\BIBentrySTDinterwordspacing

\bibitem{dpc++}
\BIBentryALTinterwordspacing
{Intel Corporation}, \emph{{Intel oneAPI Data Parallel C++}}, 2020. [Online].
  Available: \url{https://software.intel.com/en-us/oneapi/dpc-compiler}
\BIBentrySTDinterwordspacing

\bibitem{rose1987c}
J.~Rose, ``C*: An extended c language for data parallel programming,'' in
  \emph{Proceedings of the Second International Conference on Supercomputing},
  1987.

\bibitem{leissa2012extending}
R.~Lei{\ss}a, S.~Hack, and I.~Wald, ``{Extending a C-like language for portable
  SIMD programming},'' \emph{ACM SIGPLAN Notices}, vol.~47, no.~8, pp. 65--74,
  2012.

\bibitem{lattner2020mlir}
C.~Lattner, M.~Amini, U.~Bondhugula, A.~Cohen, A.~Davis, J.~Pienaar, R.~Riddle,
  T.~Shpeisman, N.~Vasilache, and O.~Zinenko, ``{MLIR: A Compiler
  Infrastructure for the End of Moore's Law},'' 2020.

\bibitem{mlirvector}
\BIBentryALTinterwordspacing
{LLVM Community}, \emph{{Multi-Level IR Compiler Framework - Vector Dialect}},
  2020. [Online]. Available: \url{https://mlir.llvm.org/docs/Dialects/Vector}
\BIBentrySTDinterwordspacing

\bibitem{truong2016latte}
L.~Truong, R.~Barik, E.~Totoni, H.~Liu, C.~Markley, A.~Fox, and T.~Shpeisman,
  ``Latte: a language, compiler, and runtime for elegant and efficient deep
  neural networks,'' in \emph{Proceedings of the 37th ACM SIGPLAN Conference on
  Programming Language Design and Implementation}, 2016, pp. 209--223.

\bibitem{rotem2018glow}
N.~Rotem, J.~Fix, S.~Abdulrasool, G.~Catron, S.~Deng, R.~Dzhabarov, N.~Gibson,
  J.~Hegeman, M.~Lele, R.~Levenstein \emph{et~al.}, ``Glow: Graph lowering
  compiler techniques for neural networks,'' \emph{arXiv preprint
  arXiv:1805.00907}, 2018.

\bibitem{chiw2012diderot}
C.~Chiw, G.~Kindlmann, J.~Reppy, L.~Samuels, and N.~Seltzer, ``Diderot: a
  parallel {DSL} for image analysis and visualization,'' in \emph{Proceedings
  of the 33rd ACM SIGPLAN conference on Programming Language Design and
  Implementation}, 2012, pp. 111--120.

\bibitem{fuentes2019lock}
J.~Fuentes, W.-Y. Chen, G.-Y. Lueh, and I.~D. Scherson, ``A lock-free skiplist
  for integrated graphics processing units,'' in \emph{2019 IEEE International
  Parallel and Distributed Processing Symposium Workshops (IPDPSW)}.\hskip 1em
  plus 0.5em minus 0.4em\relax IEEE, 2019, pp. 36--46.

\bibitem{10.1007/978-3-030-43229-4_33}
J.~Fuentes, W.-y. Chen, G.-y. Lueh, A.~Garza, and I.~D. Scherson, ``{SIMD-node
  Transformations for Non-blocking Data Structures},'' in \emph{Parallel
  Processing and Applied Mathematics}.\hskip 1em plus 0.5em minus 0.4em\relax
  Cham: Springer International Publishing, 2020, pp. 385--395.

\bibitem{chen2018register}
W.-Y. Chen, G.-Y. Lueh, P.~Ashar, K.~Chen, and B.~Cheng, ``Register allocation
  for {Intel} processor graphics,'' in \emph{Proceedings of the 2018
  International Symposium on Code Generation and Optimization}, 2018, pp.
  352--364.

\bibitem{coutinho2011divergence}
B.~Coutinho, D.~Sampaio, F.~M.~Q. Pereira, and W.~Meira~Jr, ``Divergence
  analysis and optimizations,'' in \emph{2011 International Conference on
  Parallel Architectures and Compilation Techniques}.\hskip 1em plus 0.5em
  minus 0.4em\relax IEEE, 2011, pp. 320--329.

\bibitem{cuda-warp}
\BIBentryALTinterwordspacing
{Lin, Yuan and Grover, Vinod}, \emph{{Using CUDA Warp-Level Primitives}}, 2018.
  [Online]. Available:
  \url{https://devblogs.nvidia.com/using-cuda-warp-level-primitives/}
\BIBentrySTDinterwordspacing

\bibitem{opencl-subgroup}
\BIBentryALTinterwordspacing
{Khronos OpenCL Working Group}, \emph{{The OpenCL Extension Specification}},
  2018. [Online]. Available:
  \url{https://www.khronos.org/registry/OpenCL/sdk/2.0/docs/man/xhtml/cl_khr_subgroups.html}
\BIBentrySTDinterwordspacing

\bibitem{10.1007/978-3-642-54807-9_8}
J.~Anantpur and G.~R., ``Taming control divergence in gpus through control flow
  linearization,'' in \emph{Compiler Construction}, A.~Cohen, Ed.\hskip 1em
  plus 0.5em minus 0.4em\relax Berlin, Heidelberg: Springer Berlin Heidelberg,
  2014, pp. 133--153.

\bibitem{han2011reducing}
T.~D. Han and T.~S. Abdelrahman, ``Reducing branch divergence in gpu
  programs,'' in \emph{Proceedings of the Fourth Workshop on General Purpose
  Processing on Graphics Processing Units}, 2011, pp. 1--8.

\bibitem{chandrasekhar2019igc}
A.~Chandrasekhar, G.~Chen, P.-Y. Chen, W.-Y. Chen, J.~Gu, P.~Guo, S.~H.~P.
  Kumar, G.-Y. Lueh, P.~Mistry, W.~Pan \emph{et~al.}, ``{IGC: the open source
  Intel Graphics Compiler},'' in \emph{2019 IEEE/ACM International Symposium on
  Code Generation and Optimization (CGO)}.\hskip 1em plus 0.5em minus
  0.4em\relax IEEE, 2019, pp. 254--265.

\bibitem{lattner2008llvm}
C.~Lattner, ``{LLVM and Clang}: Next generation compiler technology,'' in
  \emph{The BSD conference}, vol.~5, 2008.

\bibitem{opencl-sdk}
\BIBentryALTinterwordspacing
{Intel Corporation}, \emph{{Intel SDK for OpenCL Applications}}, 2019.
  [Online]. Available:
  \url{https://software.intel.com/en-us/opencl-sdk/training#codesamples}
\BIBentrySTDinterwordspacing

\bibitem{lee1996bitonic}
J.-D. Lee and K.~E. Batcher, ``A bitonic sorting network with simpler flip
  interconnections,'' in \emph{Proceedings Second International Symposium on
  Parallel Architectures, Algorithms, and Networks (I-SPAN'96)}.\hskip 1em plus
  0.5em minus 0.4em\relax IEEE, 1996, pp. 104--109.

\bibitem{alsabti1997efficient}
K.~Alsabti, S.~Ranka, and V.~Singh, ``An efficient k-means clustering
  algorithm,'' 1997.

\bibitem{harris2013efficient}
\BIBentryALTinterwordspacing
M.~Harris, \emph{An efficient matrix transpose in CUDA C/C++}, 2013. [Online].
  Available:
  \url{https://devblogs.nvidia.com/efficient-matrix-transpose-cuda-cc}
\BIBentrySTDinterwordspacing

\bibitem{sgemm2015ocl}
\BIBentryALTinterwordspacing
L.~Kong and R.~Ioffe, \emph{{SGEMM for Intel® Processor Graphics}}, 2015.
  [Online]. Available:
  \url{https://software.intel.com/en-us/articles/sgemm-for-intel-processor-graphics}
\BIBentrySTDinterwordspacing

\bibitem{blelloch1989scans}
G.~E. Blelloch, ``Scans as primitive parallel operations,'' \emph{IEEE
  Transactions on computers}, vol.~38, no.~11, pp. 1526--1538, 1989.

\bibitem{dpcpp-esimd}
{Intel Corporation}, ``{Explicit SIMD Programming Extension for DPC++},''
  \url{https://github.com/intel/llvm/blob/sycl/sycl/doc/extensions/ExplicitSIMD/dpcpp-explicit-simd.md},
  2020.

\bibitem{ispc-gen}
------, ``{ISPC for Gen},'' \url{https://ispc.github.io/ispc_for_gen.html},
  2020.

\end{thebibliography}
\balance
\end{document}